\documentclass[12pt]{article}
\usepackage{latexsym}
\usepackage{amsmath}

\usepackage{cite}

\date{}
\begin{document}
 \begin{center}
{\Large \textbf{On a class of second-order PDEs\\[2mm]
                admitting partner symmetries}}

\vspace{5mm}{\large\textbf{M. B. Sheftel$^{1,2}$ and A. A.
Malykh$^2$}}
\end{center}
 \vspace{2mm} $^1$ Department of Physics, Bo\u{g}azi\c{c}i
University 34342 Bebek, Istanbul, Turkey  \\
 $^2$ Department of Higher Mathematics, North Western State Technical\\
$\phantom{^1 }$ University, Millionnaya St. 5, 191186, St.
Petersburg, Russia \\
 $^3$ Department of Numerical Modelling, Russian State
Hydrometeorlogical
 \\ $\phantom{^1 }$  University, Malookhtinsky pr. 98, 195196 St.
Petersburg, Russia
 \vspace{1mm}
 \\ $\phantom{^1 }$ E-mail: mikhail.sheftel@boun.edu.tr,
            andrei-malykh@mail.ru
\begin{abstract}\noindent
 Recently we have demonstrated how to use partner symmetries for
obtaining noninvariant solutions of heavenly equations of
Pleba\~nski that govern heavenly gravitational metrics. In this
paper, we present a class of scalar second-order PDEs with four
variables, that possess partner symmetries and contain only second
derivatives of the unknown. We present a general form of such a
PDE together with recursion relations between partner symmetries.
This general PDE is transformed to several simplest canonical
forms containing the two heavenly equations of Pleba\~nski among
them and two other nonlinear equations which we call mixed
heavenly equation and asymmetric heavenly equation. On an example
of the mixed heavenly equation, we show how to use partner
symmetries for obtaining noninvariant solutions of PDEs by a lift
from invariant solutions. Finally, we present Ricci-flat self-dual
metrics governed by solutions of the mixed heavenly equation and
its Legendre transform.
\end{abstract}

\section{Introduction}

In his paper \cite{pleb}, Pleba\~nski introduced heavenly
equations for a single potential generating (anti-)self-dual
heavenly metrics which satisfy complex vacuum Einstein equations.
Two real cross sections of these complex metrics, K\"ahler metrics
with Euclidean or ultra-hyperbolic signature, are generated by the
elliptic and hyperbolic complex Monge-Amp\`ere equation ($CMA$)
respectively, the particular cases of the first heavenly equation.
Solutions of $CMA$ play an important role in the theory of
gravitational instantons \cite{egh}, where all gravitational
metrics obtained so far, apart from the ones that we obtained
lately \cite{mns,lift}, have Killing vectors, i.e. admit
continuous symmetries. This implies symmetry reduction in the
number of independent variables in metric components \cite{ibr},
so these metrics actually live on manifolds of dimensions less
than four.

Recently we introduced the concept of partner symmetries and
applied them to obtain noninvariant solutions of the complex
Monge-Amp\`ere equation and the second heavenly equation of
Pleba\~nski \cite{mns,mnsh,lift,shma}. Only such solutions could
generate heavenly metrics with no Killing vectors, so that the
metric components would depend on all four independent variables.
Partner symmetries constitute a certain type of nonlocal
symmetries and solutions invariant with respect to these nonlocal
symmetries generically are solutions noninvariant in the usual
sense, i.e. they depend on all four variables, so that no symmetry
reduction in the number of independent variables occurs. The idea
of using the invariance under nonlocal symmetries in order to get
noninvariant solutions, suggested first by Dunajski and Mason
\cite{dunman,duma}, clarified for us the meaning of differential
constraints, which we used earlier in \cite{ms,mash} to derive
non-invariant solutions of $CMA$. Since the partner symmetries and
their use for lifting invariant solutions to noninvariant ones
\cite{lift,shma} proved to be an appropriate tool for constructing
noninvariant solutions of partial differential equations (PDEs)
and a PDE taken at random would not admit partner symmetries, the
natural question arises of how general this method of obtaining
noninvariant solutions can be, or, in other words, what is the
general form of equations that possess partner symmetries?

To give a partial answer to this question, in this paper we
present some results on a classification of the second-order PDEs
of the general form
\begin{equation}\label{eq}
F(u_{tt},u_{tx},u_{ty},u_{tz},u_{xx},u_{xy},u_{xz},u_{yy},u_{yz},u_{zz},
  u_t,u_x,u_y,u_z,u,t,x,y,z) = 0
\end{equation}
that possess partner symmetries. Here $u$ is the unknown that
depends on the four independent variables $t,x,y,z$ and the
subscripts denote partial derivatives of $u$, e.g.
$u_{tt}=\partial^2 u/{\partial t}^2$, $u_{tx}=\partial^2
u/{\partial t\partial x}$ \dots Although we have derived a
complete set of equations for $F$ such that equation (\ref{eq})
admits partner symmetries, we are currently able to give a general
solution to these equations only for $F$ that depends only on
second derivatives of $u$. Thus, we obtain a classification of
PDEs of the form
\begin{equation}\label{eq2}
  F(u_{tt},u_{tx},u_{ty},u_{tz},u_{xx},u_{xy},u_{xz},u_{yy},u_{yz},u_{zz}) = 0
\end{equation}
that possess partner symmetries.

Our definition of partner symmetries requires the following two
conditions to be satisfied:
\begin{enumerate}
\item
 The symmetry condition for PDE (\ref{eq}) (determining
 equation for symmetries) has the form of a two-dimensional
 divergence, that implies the existence of a single potential for each
 symmetry.
\item
 The potential of each symmetry is itself a symmetry of (\ref{eq}), i.e. a
partner symmetry for the original symmetry.
\end{enumerate}
The first condition is satisfied in two steps: at first we require
the symmetry condition to have the form of a four-dimensional
divergence and then reduce this to a two-dimensional divergence by
imposing additional constraints on $F$. We note that it were also
possible to use a four-dimensional divergence form by introducing
several potentials, as was shown, for example, by Bluman and Kumei
\cite{blum}, which would probably modify our concept of partner
symmetries. This work is currently in progress.

In section \ref{sec:div}, we derive the restriction on the form of
equation (\ref{eq}) implied by the requirement that the symmetry
condition should have the form of a four-dimensional divergence:
the left-hand side of the equation (\ref{eq}) itself should be a
four-dimensional divergence, so that (\ref{eq}) becomes a
conservation law.

In section \ref{sec:div2}, we derive further conditions on $F$
under which the four-dimensional divergence form of the symmetry
condition is reduced to a two-dimensional divergence form which
implies the existence of a single potential for each symmetry of
(\ref{eq}).

In section \ref{sec:partner}, we require that the potential of a
symmetry should itself be a symmetry of the equation (\ref{eq})
and obtain the final set of equations for $F$. The definition of
the symmetry potential then becomes a recursion relation for
symmetries which generically maps any local symmetry into a
certain nonlocal symmetry. To have an explicit form of this
recursion relation, we still need a solution of the equations for
$F$. We note that our \emph{symmetry potential} is completely
different from \emph{potential symmetries} of Bluman and Kumei
\cite{blum}, where potentials are introduced not for symmetries
but for PDEs, set in a divergence form, and symmetries are allowed
to depend on these potential variables.

In section \ref{sec:solution}, we attempt to solve the set of
equations for $F$. The solution process in full generality turns
out to be too lengthy and suggests many cases and subcases to be
considered. Therefore, here we restrict ourselves to the case
where $F$ in (\ref{eq}) depends only on the second derivatives of
$u$ and the equation takes the form (\ref{eq2}). Then we obtain a
general solution for $F$ in the left-hand side of (\ref{eq2}),
that is, a general form of the second-order PDE with four
variables containing only second derivatives of $u$ that possesses
partner symmetries, up to a change of notation for independent
variables. We also obtain a recursion relation between partner
symmetries in an explicit form.

In section \ref{sec:canon}, we present a complete set of canonical
forms, to which the general PDE with partner symmetries can be
transformed by point and Legendre transformations, together with
recursions for symmetries of these canonical equations. Among
these canonical forms we find the first and second equations of
Pleba\~nski and two other nonlinear equations which we call mixed
heavenly equation and asymmetric heavenly equation. The mixed
heavenly equation turns out to be related by a partial Legendre
transformation to Husain's heavenly equation
\cite{husain,ppg,plebprz}, which is an alternative form of the
self-dual gravity equation related to the chiral model approach to
self-dual gravity.

In section \ref{sec:lift}, we demonstrate an application of
partner symmetries for finding noninvariant solutions of PDEs on
an example of the mixed heavenly equation. We choose both
symmetries in the recursion relations as translational symmetries,
with the recursions becoming differential constraints, and then
show that Legendre transformation in two variables of both the
equation and two differential constraints leads to a set of three
linear equations with constant coefficients. One of these
equations depends only on three variables, containing the fourth
variable merely as a parameter, and coincides with the Legendre
transform of the translational symmetry reduction of the mixed
heavenly equation, but expressed in new variables. Two other
linear equations provide a lift of any solution of this equation,
which is an invariant solution to the mixed heavenly equation, to
a noninvariant solution that depends on all four variables. We
present explicitly a linear combination of exponential solutions
and a polynomial solution as examples of such solutions.

In section \ref{sec:metrics}, we obtain Ricci-flat \textit{mixed
heavenly metric} in the self-dual gravity, governed by solutions
of the mixed heavenly equation, by using a one-dimensional
Legendre transformation of Husain's heavenly metric with a
subsequent symmetrization of the transformed metric. Then we apply
the linearizing Legendre transformation from section
\ref{sec:lift} to the mixed heavenly metric to obtain the
Ricci-flat self-dual metric with a potential satisfying the
Legendre transformed mixed heavenly equation. We are now able to
use our solutions of the latter equation, that are given in
section \ref{sec:lift}, in the obtained metric or any other
solutions of the above-mentioned three linear PDEs with constant
coefficients. In this way we arrive at an explicit form of a
Ricci-flat self-dual metric with components generically depending
on all four independent variables which, as a consequence, will
admit no continuous symmetries.

We have to mention also that a different classification of
integrable three- and four-dimensional PDEs, that contain only
second derivatives of the unknown, was given by Ferapontov et. al.
in \cite{fer,ferap}. In this approach integrability is understood
as the existence of sufficiently many hydrodynamic reductions,
which is a requirement completely different from the existence of
partner symmetries and therefore the results are also completely
different.

\section{Divergence form of symmetry condition}
\setcounter{equation}{0}
 \label{sec:div}

Let $\varphi$ be a symmetry characteristic \cite{olv} of
(\ref{eq}). Then the symmetry condition for symmetries $\varphi$
admitted by (\ref{eq}) is determined by vanishing of the Fr\'echet
derivative of $F$ on solutions of (\ref{eq})
\begin{eqnarray}
 \hat A(\varphi)  &\equiv & F_u\varphi + F_{u_t}\varphi_t +
F_{u_x}\varphi_x +
  F_{u_y}\varphi_y + F_{u_z}\varphi_z + F_{u_{tt}}\varphi_{tt} +
  F_{u_{tx}}\varphi_{tx} \nonumber
 \\ & & \mbox{} + F_{u_{ty}}\varphi_{ty} + F_{u_{tz}}\varphi_{tz} + F_{u_{xx}}\varphi_{xx} +
 F_{u_{xy}}\varphi_{xy} + F_{u_{xz}}\varphi_{xz} \nonumber
 \\ & & \mbox{} + F_{u_{yy}}\varphi_{yy}
 + F_{u_{yz}}\varphi_{yz} + F_{u_{zz}}\varphi_{zz} = 0,
 \label{DE}
\end{eqnarray}
where $\varphi_t = D_t\varphi$, $\varphi_x = D_x\varphi, \dots$
and $D_t, D_x, \dots$ denote operators of total derivatives with
respect to $t, x, \dots$, e.g.
\begin{eqnarray*}
& & D_t f = \partial f/\partial t + u_t\partial f/\partial u
 + u_{tt}\partial f/{\partial u_t} + u_{xt}\partial f/{\partial u_x}
 + u_{yt}\partial f/{\partial u_y}
 \\ & & \mbox{} + u_{zt}\partial f/{\partial u_z} + u_{ttt}\partial f/{\partial u_{tt}}
 + u_{txt}\partial f/{\partial u_{tx}} + u_{tyt}\partial f/{\partial u_{ty}} + \dots
 \end{eqnarray*}
After collecting all terms that can be written as total
derivatives, the symmetry condition (\ref{DE}) becomes
\begin{equation}\label{by_parts}
  D_t(M) + D_x(N) + D_y(L) + D_z(K) + E_u(F)\varphi = 0,
\end{equation}
where $E_u(F)$ denotes the Euler-Lagrange operator \cite{olv}
applied to $F$
\begin{eqnarray}
 & & E_u(F) = D_t^2(F_{u_{tt}}) + D_x^2(F_{u_{xx}}) +
  D_y^2(F_{u_{yy}}) + D_z^2(F_{u_{zz}}) + D_tD_x(F_{u_{tx}})
  \nonumber
 \\ & & \mbox{} + D_tD_y(F_{u_{ty}}) +
 D_tD_z(F_{u_{tz}}) + D_xD_y(F_{u_{xy}}) + D_xD_z(F_{u_{xz}})
 \nonumber
 \\ & & \mbox{} + D_yD_z(F_{u_{yz}}) - D_t(F_{u_t}) - D_x(F_{u_x}) - D_y(F_{u_y}) - D_z(F_{u_z})+ F_u
 \label{Euler}
\end{eqnarray}
and $M, N, L, K$ are defined by
\begin{eqnarray}
 & & M = F_{u_{tt}}\varphi_t + \frac{1}{2}\,F_{u_{tx}}\varphi_x +
 \frac{1}{2}\,F_{u_{ty}}\varphi_y +
 \frac{1}{2}\,F_{u_{tz}}\varphi_z \nonumber
 \\ & & \mbox{} + \left[F_{u_t} - D_t(F_{u_{tt}}) - \frac{1}{2}\,D_x(F_{u_{tx}})
 - \frac{1}{2}\,D_y(F_{u_{ty}}) -
 \frac{1}{2}\,D_z(F_{u_{tz}})\right]\varphi, \nonumber
 \\ & & N = F_{u_{xx}}\varphi_x + \frac{1}{2}\,F_{u_{tx}}\varphi_t +
 \frac{1}{2}\,F_{u_{xy}}\varphi_y +
 \frac{1}{2}\,F_{u_{xz}}\varphi_z \nonumber
 \\ & & \mbox{} + \left[F_{u_x} - D_x(F_{u_{xx}}) - \frac{1}{2}\,D_t(F_{u_{tx}})
 - \frac{1}{2}\,D_y(F_{u_{xy}}) -
 \frac{1}{2}\,D_z(F_{u_{xz}})\right]\varphi, \nonumber
 \\ & & L = F_{u_{yy}}\varphi_y + \frac{1}{2}\,F_{u_{ty}}\varphi_t +
 \frac{1}{2}\,F_{u_{xy}}\varphi_x +
 \frac{1}{2}\,F_{u_{yz}}\varphi_z
  \label{div}
 \\ & & \mbox{} + \left[F_{u_y} - D_y(F_{u_{yy}}) - \frac{1}{2}\,D_t(F_{u_{ty}})
 - \frac{1}{2}\,D_x(F_{u_{xy}}) -
 \frac{1}{2}\,D_z(F_{u_{yz}})\right]\varphi, \nonumber
 \\ & & K = F_{u_{zz}}\varphi_z + \frac{1}{2}\,F_{u_{tz}}\varphi_t +
 \frac{1}{2}\,F_{u_{xz}}\varphi_x +
 \frac{1}{2}\,F_{u_{yz}}\varphi_y \nonumber
 \\ & & \mbox{} + \left[F_{u_z} - D_z(F_{u_{zz}}) - \frac{1}{2}\,D_t(F_{u_{tz}})
 - \frac{1}{2}\,D_x(F_{u_{xz}}) -
 \frac{1}{2}\,D_y(F_{u_{yz}})\right]\varphi . \nonumber
\end{eqnarray}
The determining equation, transformed to the form
(\ref{by_parts}), takes the divergence form on solutions of
(\ref{eq})
\begin{equation}\label{div4}
  D_t(M) + D_x(N) + D_y(L) + D_z(K) = 0,
\end{equation}
if and only if the Euler-Lagrange equation
\begin{equation}\label{E-L}
  E_u(F) = 0
\end{equation}
is identically satisfied on solutions of $F = 0$, which is
equivalent to the\newline $4$-divergence form of the equation
(\ref{eq}) itself \cite{olv}:
\begin{equation}\label{Ediv}
  F \equiv D_t(P) + D_x(Q) + D_y(R) + D_z(S) = 0,
\end{equation}
where $P, Q, R, S$ depend on the same set of variables as $F$ in
(\ref{eq}).

\section{Two-dimensional divergence form of the\\ symmetry condition}
\setcounter{equation}{0}
 \label{sec:div2}

In order to introduce a unique potential as a consequence of a
symmetry condition, we have to convert the four-dimensional
divergence in the left-hand side of the symmetry condition
(\ref{div4}) into a two-dimensional divergence, say, in the
variables $t$ and $x$. To do this, we present $L$ and $K$ as the
sum of total derivatives in $t$ and $x$ plus remaining terms which
cannot be given in this form:
\begin{eqnarray}
& & L = D_t\left(\frac{1}{2} F_{u_{ty}}\varphi\right) +
D_x\left(\frac{1}{2} F_{u_{xy}}\varphi\right) +
F_{u_{yy}}\varphi_y + \frac{1}{2} F_{u_{yz}}\varphi_z \nonumber
\\ & & \mbox{} + \left[F_{u_y} - D_y(F_{u_{yy}}) - D_t(F_{u_{ty}}) - D_x(F_{u_{xy}})
- \frac{1}{2} D_z(F_{u_{zy}})\right]\varphi ,
 \label{LK_tx}
 \\ & & K = D_t\left(\frac{1}{2} F_{u_{tz}}\varphi\right) +
D_x\left(\frac{1}{2} F_{u_{xz}}\varphi\right) +
F_{u_{zz}}\varphi_z + \frac{1}{2} F_{u_{yz}}\varphi_y \nonumber
\\ & & \mbox{} + \left[F_{u_z} - D_z(F_{u_{zz}}) - D_t(F_{u_{tz}}) - D_x(F_{u_{xz}})
- \frac{1}{2} D_y(F_{u_{yz}})\right]\varphi. \nonumber
\end{eqnarray}
Using (\ref{LK_tx}) in (\ref{div4}) together with the definitions
(\ref{div}) and collecting terms with the total derivatives with
respect to $t$ and $x$, we convert (\ref{div4}) to the form
\begin{eqnarray}
& & D_t(\bar M) - D_x(\bar N) + F_{u_{yy}}\varphi_{yy} +
F_{u_{yz}}\varphi_{yz} + F_{u_{zz}}\varphi_{zz} \nonumber
 \\ & & \mbox{} + \left[F_{u_y} - D_t(F_{u_{ty}}) - D_x(F_{u_{xy}})\right]
 \varphi_y + \left[F_{u_z} - D_t(F_{u_{tz}}) - D_x(F_{u_{xz}})\right]
 \varphi_z \nonumber
 \\ & & \mbox{} + \left\{D_y\left[F_{u_y} - D_y(F_{u_{yy}}) - D_t(F_{u_{ty}})
 - D_x(F_{u_{xy}})\right]\right.
  \label{div2+}
 \\ & & \left. \mbox{} + D_z\left[F_{u_z} - D_z(F_{u_{zz}}) - D_t(F_{u_{tz}})
 - D_x(F_{u_{xz}})\right] - D_yD_z(F_{u_{yz}})\right\}\varphi = 0,
 \nonumber
\end{eqnarray}
where $\bar M$ and $\bar N$ are defined by
\begin{eqnarray}
& & \bar M = F_{u_{tt}}\varphi_t + \frac{1}{2}F_{u_{tx}}\varphi_x
+ F_{u_{ty}}\varphi_y + F_{u_{tz}}\varphi_z \nonumber
\\ & & \mbox{} + \left[F_{u_t} - D_t(F_{u_{tt}}) -
\frac{1}{2}D_x(F_{u_{tx}})\right]\varphi ,
 \label{barMN}
 \\ & & \bar N = -\left\{F_{u_{xx}}\varphi_x +
 \frac{1}{2}F_{u_{tx}}\varphi_t
+ F_{u_{xy}}\varphi_y + F_{u_{xz}}\varphi_z\right. \nonumber
\\ & & \left.\mbox{} + \left[F_{u_x} - D_x(F_{u_{xx}}) -
\frac{1}{2}D_t(F_{u_{tx}})\right]\varphi\right\} . \nonumber
\end{eqnarray}
It is clear that in order to have the symmetry condition
(\ref{div2+}) to be a two-dimensional divergence in the variables
$t$ and $x$ the coefficients of all the terms not included in the
total derivatives $D_t$ and $D_x$ should vanish on solutions of
(\ref{eq}):
\begin{equation}\label{Fij}
  F_{u_{yy}} =   F_{u_{yz}} = F_{u_{zz}} = 0 ,
\end{equation}
\begin{equation}\label{Fyz}
  F_{u_y} - D_t(F_{u_{ty}}) - D_x(F_{u_{xy}}) = 0, \quad
  F_{u_z} - D_t(F_{u_{tz}}) - D_x(F_{u_{xz}}) = 0
\end{equation}
whereas, as a consequence of (\ref{Fij}) and (\ref{Fyz}), the
coefficient of $\varphi$ in (\ref{div2+}) vanishes identically and
the symmetry condition (\ref{div2+}) becomes
\begin{equation}\label{div2}
  D_t(\bar M) = D_x(\bar N).
\end{equation}
Note that the symmetry condition and therefore all the equations
(\ref{div2+}), (\ref{Fij}), (\ref{Fyz}), and (\ref{div2}) should
be satisfied not identically but only on solutions of the original
PDE (\ref{eq}) and hence they should be (differential)
consequences of $F = 0$.

Condition (\ref{div2}) is equivalent to the local existence of the
potential $\psi$ defined by
\begin{equation}\label{psi}
  \psi_t = \hat N = \bar N + \Lambda_t,\qquad \psi_x = \hat M = \bar M + \Lambda_x
  , \qquad \Lambda = \omega\varphi ,
\end{equation}
where $\omega$ may depend on $t,x,y,z,u$ and the first and second
derivatives of $u$. Here the terms with the derivatives of
$\Lambda$ are added in order to have the most general definition
of the potential $\psi$. Now, the symmetry condition (\ref{div2+})
can be written as
\begin{equation}\label{div_2}
  D_t(\hat M) = D_x(\hat N)
\end{equation}
on solutions of $F = 0$.

\section{Existence conditions for partner\\ symmetries}
\setcounter{equation}{0}
 \label{sec:partner}

Our second requirement is that the potential $\psi$ should also be
a symmetry of the PDE (\ref{eq}), i.e. a partner symmetry for the
original symmetry $\varphi$, so that (\ref{psi}) becomes a
recursion relation for symmetries. Then the symmetry condition in
the two-dimensional divergence form (\ref{div_2}) with $\varphi$
replaced by $\psi$, defined by (\ref{psi}), should be satisfied on
solutions of the equation (\ref{eq})
\begin{equation}
 D_t\left(\tilde{M}\right) = D_x\left(\tilde{N}\right) + \hat F ,
 \label{DEpsi}
\end{equation}
where $\tilde{M}$ and $\tilde{N}$ are obtained from $\bar M$ and
$\bar N$ respectively by replacing $\varphi$ with $\psi$ in
(\ref{barMN})
\begin{eqnarray}
& & \tilde M = F_{u_{tt}}\psi_t + \frac{1}{2}F_{u_{tx}}\psi_x +
F_{u_{ty}}\psi_y + F_{u_{tz}}\psi_z \nonumber
\\ & & \mbox{} + \left[F_{u_t} - D_t(F_{u_{tt}}) -
\frac{1}{2}D_x(F_{u_{tx}})\right]\psi ,
 \label{tildeMN}
 \\ & & \tilde N = -\left\{F_{u_{xx}}\psi_x +
 \frac{1}{2}F_{u_{tx}}\psi_t
+ F_{u_{xy}}\psi_y + F_{u_{xz}}\psi_z\right. \nonumber
\\ & & \left.\mbox{} + \left[F_{u_x} - D_x(F_{u_{xx}}) -
\frac{1}{2}D_t(F_{u_{tx}})\right]\psi\right\} . \nonumber
\end{eqnarray}
The term $\hat F$ has the form
\begin{equation}\label{hatF}
  \hat F = \mu D_t(F) + \nu D_x(F) + \rho D_y(F) + \lambda D_z(F)
  + \sigma F
\end{equation}
and it accounts for the fact that equation (\ref{DEpsi}) should be
satisfied only on solutions of (\ref{eq}) (a consequence of
proposition 2.10 in \cite{olv}, similar to formula (2.26)
therein). Terms with $\psi_y$, $\psi_z$, and $\psi$ in
(\ref{DEpsi}) cannot be balanced by any other terms and therefore
they should vanish separately on solutions of (\ref{eq}) yielding
\begin{equation}\label{psi_yz}
  D_t(F_{u_{ty}}) + D_x(F_{u_{xy}}) = \hat F^y,\quad D_t(F_{u_{tz}}) + D_x(F_{u_{xz}}) =
  \hat F^z
\end{equation}
and
\begin{equation}\label{psi_}
  D_t(F_{u_t}) +   D_x(F_{u_x}) - D_t^2(F_{u_{tt}}) -
  D_x^2(F_{u_{xx}}) - D_tD_x(F_{u_{tx}}) = \hat F
\end{equation}
respectively, where the terms $\hat F^y$ and $\hat F^z$ are of the
same form (\ref{hatF}) but with different coefficients $\mu, \nu,
\rho, \lambda, \sigma$. Equations (\ref{psi_yz}) together with
(\ref{Fyz}) and equation (\ref{psi_}) together with (\ref{E-L})
imply
\begin{equation}\label{F_u}
  F_{u_y} = 0,\quad F_{u_z} = 0,\quad F_u = 0.
\end{equation}
In all other terms in (\ref{DEpsi}), we replace $\psi_t$ and
$\psi_x$ by the expressions (\ref{psi}). We note that, due to the
definition (\ref{psi}) of the potential $\psi$ and its consequence
(\ref{div_2}) (equivalent to (\ref{div2})), $\varphi$ satisfies
the symmetry condition (\ref{DE}), which cancels all the terms
proportional to $\omega$ in (\ref{DEpsi}). All other terms in
(\ref{DEpsi}) with second derivatives of $\varphi$ are cancelled
identically. The remaining terms are proportional to $\varphi_t$,
$\varphi_x$, $\varphi_y$, $\varphi_z$, and $\varphi$, so that
these five groups of terms should vanish separately on solutions
of the equation $F = 0$ to give the following five equations
respectively:
\begin{eqnarray}
& & D_x(F_{u_{tt}}F_{u_{xx}}) - \frac{1}{4}\,D_x(F_{u_{tx}}^2) +
 F_{u_{xy}}D_y(F_{u_{tt}}) + F_{u_{xz}}D_z(F_{u_{tt}}) \nonumber
 \\ & & \mbox{} - \frac{1}{2}\big[F_{u_{ty}}D_y(F_{u_{tx}})
 + F_{u_{tz}}D_z(F_{u_{tx}})\big] + 2F_{u_{tt}}D_t(\omega) +
 F_{u_{tx}}D_x(\omega) \nonumber
  \\ & & \mbox{} + F_{u_{ty}}D_y(\omega) +  F_{u_{tz}}D_z(\omega) = \hat
  F_1 ,
\label{fi_t}
\end{eqnarray}
\begin{eqnarray}
& & -\Big\{D_t(F_{u_{tt}}F_{u_{xx}}) -
\frac{1}{4}\,D_t(F_{u_{tx}}^2) +
 F_{u_{ty}}D_y(F_{u_{xx}}) + F_{u_{tz}}D_z(F_{u_{xx}}) \nonumber
 \\ & & \mbox{} - \frac{1}{2}\big[F_{u_{xy}}D_y(F_{u_{tx}})
 + F_{u_{xz}}D_z(F_{u_{tx}})\big] - 2F_{u_{xx}}D_x(\omega) -
 F_{u_{tx}}D_t(\omega) \nonumber
  \\ & & \mbox{} - F_{u_{xy}}D_y(\omega) -
  F_{u_{xz}}D_z(\omega)\Big\} = \hat F_2 ,
\label{fi_x}
\end{eqnarray}
\begin{eqnarray}
& & D_x(F_{u_{ty}}F_{u_{xx}}) -
\frac{1}{2}\,D_x(F_{u_{tx}}F_{u_{xy}}) -
 D_t(F_{u_{xy}}F_{u_{tt}}) +
 \frac{1}{2}\,D_t(F_{u_{tx}}F_{u_{ty}})\nonumber
\\ & & \mbox{} + F_{u_{xy}}D_y(F_{u_{ty}}) +
F_{u_{xz}}D_z(F_{u_{ty}})- F_{u_{ty}}D_y(F_{u_{xy}})
 - F_{u_{tz}}D_z(F_{u_{xy}})
\nonumber
 \\ & & \mbox{} + F_{u_{ty}}D_t(\omega) +
 F_{u_{xy}}D_x(\omega) = \hat F_3 ,
 \label{fi_y}
\end{eqnarray}
\begin{eqnarray}
& & D_x(F_{u_{tz}}F_{u_{xx}}) -
\frac{1}{2}\,D_x(F_{u_{tx}}F_{u_{xz}}) -
 D_t(F_{u_{xz}}F_{u_{tt}}) +
 \frac{1}{2}\,D_t(F_{u_{tx}}F_{u_{tz}}) \nonumber
\\ & & \mbox{} + F_{u_{xy}}D_y(F_{u_{tz}}) +
F_{u_{xz}}D_z(F_{u_{tz}})- F_{u_{ty}}D_y(F_{u_{xz}})
 - F_{u_{tz}}D_z(F_{u_{xz}})
\nonumber
 \\ & & \mbox{} + F_{u_{tz}}D_t(\omega) +
 F_{u_{xz}}D_x(\omega) = \hat F_4 ,
 \label{fi_z}
\end{eqnarray}
\begin{eqnarray}
 & & F_{u_t}B - F_{u_x}A + F_{u_{tt}}D_t(B) - F_{u_{xx}}D_x(A) +
  \frac{1}{2}\,F_{u_{tx}}\big[D_x(B) - D_t(A)\big] \nonumber
 \\ & & + F_{u_{ty}}D_y(B) + F_{u_{tz}}D_z(B) - F_{u_{xy}}D_y(A) -
 F_{u_{xz}}D_z(A) + \hat A(\omega) = \hat F_0 , \nonumber \\
 \label{fi}
\end{eqnarray}
where
\begin{equation}\label{AB}
  A = D_t(F_{u_{tt}}) + \frac{1}{2}\,D_x(F_{u_{tx}}) - F_{u_t},
  \quad B = D_x(F_{u_{xx}}) + \frac{1}{2}\,D_t(F_{u_{tx}}) - F_{u_x}
\end{equation}
and $\hat A$ is the operator of the symmetry condition (\ref{DE}).
Here the terms $\hat F_i$, of the form (\ref{hatF}) but with
different coefficients, account for the fact that the equations
should be satisfied only on solutions of (\ref{eq}).

We note that in the notation (\ref{AB}) equation (\ref{psi_})
simplifies to
\begin{equation}\label{psiAB}
  D_t(A) + D_x(B) = \hat F.
\end{equation}

\section{Equations that admit partner symmetries and recursion
relation for symmetries}
 \setcounter{equation}{0}
 \label{sec:solution}

We proceed now to solve the existence conditions for partner
symmetries (\ref{psi_yz}), (\ref{fi_t}), (\ref{fi_x}),
(\ref{fi_y}), (\ref{fi_z}), (\ref{fi}) and (\ref{psiAB}) for the
unknown left-hand side $F$ of the equation (\ref{eq}) and $\omega$
in the definition (\ref{psi}) of the potential $\psi$. We split
these equations in third derivatives of $u$ obtaining
over-determined sets of equations which can be easily solved. Our
strategy is to choose the function $\omega$ and the coefficients
$\mu,\nu,\rho,\lambda,\sigma$ in the terms $\hat F$ of the form
(\ref{hatF}) in such a way as to have minimum restrictions on the
form $F$ of equation (\ref{eq}).

We start with the equations (\ref{psi_yz}) since they do not
contain $\omega$. Our strategy results in vanishing of
$\mu,\nu,\rho,\lambda$ and $\sigma$ in $\hat F^y$ and $\hat F^z$
that implies the linear dependence of $F$ on $u_{ty}, u_{xy},
u_{tz}$ and $u_{xz}$, so that the solution of the equations
(\ref{psi_yz}) has the form
\begin{eqnarray}
& &  F = a_1(y,z) (u_{ty}u_{xz} - u_{tz}u_{xy}) + a_2
(u_{tx}u_{ty} - u_{tt}u_{xy}) + a_3 (u_{ty}u_{xx} - u_{tx}u_{xy})
\nonumber
\\ & & \mbox{} + a_4 (u_{tx}u_{tz} - u_{tt}u_{xz}) + a_5 (u_{tz}u_{xx} -
u_{tx}u_{xz}) + b_1 u_{xy} + b_2 u_{ty} \nonumber
\\ & & \mbox{} + b_3 u_{xz} + b_4 u_{tz} + g_3(u_{tt},u_{tx},u_{xx},u_t,u_x,t,x,y,z),
 \label{solpsiyz}
\end{eqnarray}
where the coefficients $a_2,a_3,a_4,a_5,b_1,b_2,b_3$ and $b_4$ are
functions of $u_t,u_x$, $t,x$, $y$, $z$ that satisfy a certain
overdetermined set of partial differential equations.

 To simplify the analysis, we assume from now on that all the
coefficients in (\ref{solpsiyz}) are constants, so that all the
equations for the coefficients are identically satisfied, and
$g_3$ depends only on the second derivatives
$u_{tt},u_{tx},u_{xx}$. As a consequence, the left-hand side $F$
of our equation (\ref{eq}) depends only on second derivatives of
$u$ and it takes the form (\ref{eq2}).

With these restrictions, we substitute the expression
(\ref{solpsiyz}) for $F$ in the remaining six equations
(\ref{fi_t}), (\ref{fi_x}), (\ref{fi_y}), (\ref{fi_z}), (\ref{fi})
and (\ref{psiAB}). The resulting equations are split in third
derivatives of $u$ into over-determined sets of equations, where
we choose the function $\omega$ and the coefficients
$\mu,\nu,\rho,\lambda,\sigma$ in the terms of the form
(\ref{hatF}) in such a way as to obtain minimum restrictions on
the form of $F$. It turns out that all these six equations
determine only the form of the function $g_3$
\begin{equation}\label{g_3}
  g_3(u_{tt},u_{tx},u_{xx}) = a_6(u_{tt}u_{xx} - u_{tx}^2) +
  b_5 u_{tt} + 2b_6 u_{tx} + b_7 u_{xx} + b_0 ,
\end{equation}
so that the equation (\ref{eq}) becomes
\begin{eqnarray}
&\hspace*{-5pt} & F = a_1(u_{ty}u_{xz} - u_{tz}u_{xy}) + a_2
(u_{tx}u_{ty} - u_{tt}u_{xy})  + a_3(u_{ty}u_{xx} -
u_{tx}u_{xy})\nonumber
\\ &\hspace*{-5pt} & \mbox{} + a_4 (u_{tx}u_{tz} - u_{tt}u_{xz})
 + a_5 (u_{tz}u_{xx} - u_{tx}u_{xz}) + a_6(u_{tt}u_{xx} - u_{tx}^2) \nonumber
\\ &\hspace*{-5pt} & \mbox{} + b_1 u_{xy} + b_2 u_{ty} + b_3 u_{xz} + b_4 u_{tz}
+ b_5 u_{tt} + 2b_6 u_{tx} + b_7 u_{xx} + b_0 = 0
 \label{Ffinal}
\end{eqnarray}
together with the following solution for $\omega$
\begin{equation}\label{omeg}
  \omega = - \frac{1}{2}\,\big(a_2u_{ty} + a_3u_{xy} + a_4u_{tz} + a_5 u_{xz}\big)+ \omega_0 ,
\end{equation}
where all the coefficients are constants. Using (\ref{omeg}) in
the equations (\ref{psi}), that define the symmetry potential
$\psi$ in terms of the symmetry $\varphi$, we obtain the recursion
relation between partner symmetries of the equation (\ref{Ffinal})
\begin{eqnarray}
 & & \psi_t = - \big(a_2u_{ty} + a_4u_{tz} - a_6 u_{tx} + b_6
 - \omega_0\big)\varphi_t \nonumber
 \\ & & \mbox{} - \big(a_3u_{ty} + a_5u_{tz} + a_6 u_{tt} +
 b_7\big)\varphi_x + \big(a_1u_{tz} + a_2u_{tt} + a_3u_{tx} -
 b_1\big)\varphi_y \nonumber
 \\ & & \mbox{} + \big(- a_1u_{ty} + a_4u_{tt} + a_5u_{tx} -
 b_3\big)\varphi_z ,
  \label{recurs}
 \\ & & \psi_x = - \big(a_2u_{xy} + a_4u_{xz} - a_6 u_{xx} - b_5\big)\varphi_t
 \nonumber
 \\ & & \mbox{} - \big(a_3u_{xy} + a_5u_{xz} + a_6 u_{tx} - b_6 - \omega_0\big)\varphi_x  \nonumber
 \\ & & \mbox{} + \big(a_1u_{xz} + a_2u_{tx} + a_3u_{xx} +
 b_2\big)\varphi_y + \big(- a_1u_{xy} + a_4u_{tx} + a_5u_{xx} +
 b_4\big)\varphi_z . \nonumber
\end{eqnarray}
Here, by construction, both $\varphi$ and $\psi$ satisfy the
symmetry condition (\ref{DE}) in the divergence form (\ref{div_2})
and (\ref{DEpsi}) respectively, on solutions of (\ref{eq2}), and
hence the transformation (\ref{recurs}) maps any symmetry
$\varphi$ of the equation (\ref{Ffinal}) again into its symmetry
$\psi$.

 \section{Canonical forms of the PDEs that admit partner symmetries}
 \setcounter{equation}{0}
 \label{sec:canon}

Due to the random choice of original variables, both the form of
the equation (\ref{Ffinal}), that admits partner symmetries, and
the recursion relation (\ref{recurs}) contain false generality.
Therefore, we present here simple canonical forms, to which the
equation (\ref{Ffinal}) can be transformed by point and Legendre
transformations, and the corresponding recursion relations for
symmetries. This will also make up for our casual choice of
variables $t$ and $x$ for the two-dimensional divergence form. The
transformations, providing the proof of the results presented
here, will be published elsewhere.
 \vspace{3mm}
 \\ \textbf{Case I: $a_1\ne 0.$}
 \\ In this case we can make $a_1 = 1$ by dividing (\ref{Ffinal})
over $a_1$. The equation (\ref{Ffinal}) can be reduced to the form
\begin{eqnarray}
& & F = u_{ty}u_{xz} - u_{tz}u_{xy} + \Gamma (u_{tt}u_{xx} -
u_{tx}^2) + A u_{tt} + B u_{xx} + C u_{tx} \nonumber
\\ & & \mbox{} + b_1 u_{xy} + b_2 u_{ty} + b_3 u_{xz} + b_4 u_{tz}
 + b_0 = 0 .
 \label{caseI}
\end{eqnarray}
Consider here the following subcases.
 \vspace{3mm}
 \\ \textbf{Subcase Ia: $\Gamma = 0.$}
 \\ Then, by the change of the unknown the equation (\ref{caseI}) becomes
\begin{equation}\label{Ia}
  u_{ty}u_{xz} - u_{tz}u_{xy} + A u_{tt} + B u_{xx} + C u_{tx} +
  D = 0.
\end{equation}
 \textbf{Subcase Ia1: $A = B = C = D = 0.$}
 \\ Then equation (\ref{Ia}) reduces to the homogeneous version of
 the first heavenly equation of Pleba\~nski
\begin{equation}\label{coef=0}
  u_{ty}u_{xz} - u_{tz}u_{xy} = 0 .
\end{equation}
 \textbf{Subcase Ia2: $A = B = C = 0,\; D\ne 0.$}
 \\ Then we can set $D = -1$ and the equation (\ref{Ia}) becomes
 the first heavenly equation of Pleba\~nski
\begin{equation}\label{heav1}
 u_{ty}u_{xz} - u_{tz}u_{xy} = 1 .
\end{equation}
In the cases Ia1 and Ia2, the recursion relation for symmetries
(\ref{recurs}) becomes
\begin{equation}\label{Ia1_2}
  \psi_t = \omega_0\varphi_t + u_{tz}\varphi_y - u_{ty}\varphi_z
,\quad \psi_x = \omega_0\varphi_x + u_{xz}\varphi_y -
u_{xy}\varphi_z.
\end{equation}
 \textbf{Subcase Ia3:  $(A,B,C) \ne (0,0,0).$}
 \\ Then we can always make $D\ne 0$. By a combination of Legendre
and point transformations the equation (\ref{Ia}) takes the
canonical form
\begin{equation}\label{Ia3}
u_{ty}u_{xz} - u_{tz}u_{xy} + u_{tt}u_{xx}-u_{tx}^2 = \varepsilon,
\end{equation}
where $\varepsilon = \pm 1$. We call (\ref{Ia3}) the \textit{mixed
heavenly equation}.

The homogeneous version of the mixed heavenly equation
(\ref{Ia3}), with $\varepsilon = 0$, can be transformed to the
first heavenly equation by an appropriate Legendre transformation.

The recursion relation (\ref{recurs}) for symmetries of equation
(\ref{Ia3}) becomes
\begin{eqnarray}\label{recursIa3}
& &  \psi_t = (u_{tx} + \omega_0)\varphi_t - u_{tt}\varphi_x +
u_{tz}\varphi_y - u_{ty}\varphi_z ,\nonumber
 \\ & & \psi_x = u_{xx}\varphi_t - (u_{tx} - \omega_0)\varphi_x + u_{xz}\varphi_y -
u_{xy}\varphi_z.
\end{eqnarray}
Recently we became aware of the relation of the mixed heavenly
equation to the Husain's heavenly equation  (at $\varepsilon = +
1$) \cite{husain,ppg} arising in the chiral model approach to
self-dual gravity
\begin{equation}\label{husain}
  v_{ty}v_{pz} - v_{tz}v_{py} + v_{tt} + \varepsilon v_{pp} = 0.
\end{equation}
Husain's equation can be obtained from the mixed heavenly equation
by the partial Legendre transformation in $x$
\begin{equation}\label{legmixhus}
 p = u_x,\quad v(t,p,y,z) = u - xu_x,\qquad x = - v_p,\quad u = v - pv_p .
\end{equation}
 We note that (\ref{husain}) could also be obtained as a canonical
equation in the subcase Ia3 of the general equation (\ref{Ffinal})
with the replacement $u\mapsto v, x\mapsto p$ with the following
choice of the coefficients in (\ref{Ffinal}): $a_1 = 1, b_5 = 1,
b_7 = \varepsilon$ and all other coefficients vanishing. Then from
(\ref{recurs}), with this change of notation, we obtain the
recursion for partner symmetries of equation (\ref{husain})
\begin{equation}\label{recurhus}
  \psi_t = \omega_0\varphi_t + v_{tz}\varphi_y - v_{ty}\varphi_z -
 \varepsilon\varphi_p,\qquad \psi_p = \varphi_t +
 \omega_0\varphi_p + v_{pz}\varphi_y - v_{py}\varphi_z .
\end{equation}
 \\ \textbf{Subcase Ib: $\Gamma \ne 0.$}
 \\ In this case, the equation (\ref{caseI}) can be transformed to
the same equation (\ref{Ia3}).
 \vspace{3mm}
 \\ \textbf{Case II: $a_1 = 0.$}
  \vspace{3mm}
 \\ \textbf{Subcase IIa: $a_2 = a_3 = a_4 = a_5 = 0,\; a_6\ne 0.$}
 \\ Then we can set $a_6 = 1$ and the equation (\ref{Ffinal}) becomes
\begin{eqnarray}
 F = u_{tt}u_{xx} - u_{tx}^2 + b_5u_{tt} + 2b_6u_{tx} +
 b_7u_{xx} \nonumber
\\ \mbox{} + b_1u_{xy} + b_2u_{ty} + b_3u_{xz} + b_4u_{tz} + b_0 =
 0.
 \label{IIa}
\end{eqnarray}
The equation (\ref{IIa}) can be transformed to the equation
\begin{equation}\label{IIaa1}
u_{tt}u_{xx} - u_{tx}^2 + b_1u_{xy} + b_2u_{ty} + b_3u_{xz} +
b_4u_{tz} = 0.
\end{equation}
We consider only the case when $(b_1,b_2)\ne (0,0)$ and
$(b_3,b_4)\ne (0,0)$. Otherwise the equation (\ref{IIaa1}) will
determine a function of less than four variables.
 \vspace{2mm}
 \\ \textbf{Case IIa1:} $b_1b_4 - b_2b_3\ne 0$.
\\ Then by a change of variables the equation (\ref{IIaa1}) takes the form of the second
heavenly equation of Pleba\~nski
\begin{equation}\label{IIa1}
u_{tt}u_{xx} - u_{tx}^2 + u_{xy} + u_{tz} = 0 .
\end{equation}
The recursion (\ref{recurs}) for symmetries of (\ref{IIa1}) takes
the form
\begin{equation}\label{recursIIa1}
\psi_t = (u_{tx} + \omega_0)\varphi_t - u_{tt}\varphi_x -
\varphi_y ,\quad \psi_x = u_{xx}\varphi_t - (u_{tx} -
\omega_0)\varphi_x + \varphi_z,
\end{equation}
which at $\omega_0=0$ coincides, up to the change $\psi\mapsto
-\psi$, with our previous result \cite{mnsh}.
 \vspace{2mm}
 \\ \textbf{Case IIa2:} $b_1b_4 - b_2b_3 = 0$.
\\ Then, by choosing a certain linear combination of $y$ and $z$,
we obtain the equation which determines a function of only three
variables.
  \vspace{3mm}
 \\ \textbf{Subcase IIb: $a_1=a_2=a_3=a_4=a_5=a_6=0.$}
 \\ Then (\ref{Ffinal}) reduces to the linear equation
\begin{eqnarray}
b_5u_{tt} + 2b_6u_{tx} + b_7u_{xx} + b_1u_{xy} + b_2u_{ty} +
b_3u_{xz} + b_4u_{tz} + b_0 = 0.
 \label{IIb}
\end{eqnarray}
The recursion (\ref{recurs}) for symmetries of (\ref{IIb}) becomes
\begin{eqnarray}
 & & \psi_t = -(b_6 - \omega_0)\varphi_t - b_7\varphi_x -
 b_1\varphi_y - b_3\varphi_z,\nonumber
 \\ & & \psi_x = b_5\varphi_t + (b_6 + \omega_0)\varphi_x +
 b_2\varphi_y + b_4\varphi_z .
 \label{linrecurs}
\end{eqnarray}
 \\ \textbf{Subcase IIc: $a_1=0,\;(a_2,a_3,a_4,a_5)\ne (0,0,0,0).$}
 \\ The equation (\ref{Ffinal}) with $a_1 = 0$ can be reduced to the canonical form
\begin{equation}\label{IIc}
  u_{tx}u_{ty} - u_{tt}u_{xy} + au_{tz} + bu_{xz} + cu_{xx} = 0 ,
\end{equation}
up to a possible change of notation for independent variables
$x\mapsto t$ and/or $z\mapsto y$. Here we can set $a=1$ by an
appropriate scaling of variables. We call this equation
\textit{asymmetric heavenly equation}. At $b=0$ it becomes the
so-called \textit{evolution form of the second heavenly equation}
\cite{plebprz,BoyPleb,FPPG}.

The recursion relation (\ref{recurs}) for symmetries of
(\ref{IIc}) takes the form
\begin{eqnarray}
 & & \psi_t = - (u_{ty} - \omega_0)\varphi_t - c\varphi_x +
 u_{tt}\varphi_y - b\varphi_z \nonumber
 \\ & & \psi_x = - u_{xy}\varphi_t + \omega_0\varphi_x +
 u_{tx}\varphi_y + a\varphi_z.
\end{eqnarray}

\section{Lift from invariant to noninvariant\\ solutions of the
         mixed heavenly equation}
 \setcounter{equation}{0}
 \label{sec:lift}

 In the method of partner symmetries, we consider a nonlocal
symmetry with the characteristic $\hat{\eta} = \tilde{\varphi} -
R\varphi$, where $\tilde{\varphi}$ is any point symmetry of our
equation and $R$ is the recursion operator determined by recursion
relations (\ref{recurs}), generating a nonlocal symmetry $\psi =
R\varphi$ from a point symmetry $\varphi$. We search for solutions
invariant with respect to a nonlocal symmetry $\hat{\eta}$,
determined by the condition $ \tilde{\varphi} - R\varphi = 0$
\cite{mns}, so that we can obtain this invariance condition by
formally replacing $\psi$ by a point symmetry $\tilde{\varphi}:$
$\psi = \tilde{\varphi}$ in the recursion relations
(\ref{recurs}). This does not mean symmetry reduction, so that
generically these solutions depend on all four variables and so
they are still noninvariant solutions in the usual sense. Contact
symmetries can also be used for $\varphi$ and/or
$\tilde{\varphi}$.

Here we demonstrate the application of partner symmetries for
obtaining noninvariant solutions of canonical PDEs and, in
particular, a \textit{lift from invariant to noninvariant
solutions.} We choose mixed heavenly equation (\ref{Ia3}) as an
example, possessing the recursion for symmetries
(\ref{recursIa3}), where we set $\omega_0 = 0$. The equation
(\ref{Ia3}) admits the obvious translational symmetry with the
generator $X =
\partial_x +\partial_z$.

Solutions, \textit{invariant} under this symmetry, have the form
$u = u(s,t,y)$, where $s = x - z$, since they do not change under
the simultaneous shift in $x$ and $z$. Then $u$ satisfies the
reduced equation
\begin{equation}\label{redeq}
  u_{ts}u_{sy} - u_{ty}u_{ss} + u_{tt}u_{ss} - u_{ts}^2 =
  \varepsilon,
\end{equation}
obtained from (\ref{Ia3}) by the symmetry reduction. Under the
Legendre transformation
\begin{equation}\label{LegIa3}
  r = u_s,\quad v(r,t,y) = u - su_s,\quad s = -v_r,\quad u = v
  - rv_r
\end{equation}
the equation (\ref{redeq}) is linearized in the form
\begin{equation}\label{linred}
  v_{tt} + \varepsilon v_{rr} - v_{ty} = 0 .
\end{equation}
By using partner symmetries, we shall show that solutions of the
linear equation (\ref{linred}), i.e. \textit{invariant solutions}
of Legendre transformed mixed heavenly equation, being written in
certain new coordinates, \textit{can be lifted up to noninvariant
solutions} of the latter equation.

 As was explained at the beginning of this section,
we formally replace $\psi$, that is generated from a point
symmetry $\varphi$ in the recursion relations (\ref{recursIa3})
(with $\omega_0 = 0$), by a point symmetry $\tilde{\varphi}$. Here
we choose both $\varphi$ and $\psi = \tilde{\varphi}$
 to be the indicated above combination of
translations in $x$ and $z$ with the characteristic $\psi =
\varphi = u_x + u_z$, so that (\ref{recursIa3}) becomes
\begin{eqnarray}
& & u_{xx} + u_{xz} = u_{tz}u_{xx} - u_{tx}u_{xz} + u_{xz}u_{yz} -
u_{xy}u_{zz} ,
 \label{redrec2}
 \\  & &  u_{tx} + u_{tz} = u_{tx}(u_{tx} + u_{tz}) - u_{tt}(u_{xx} +  u_{xz})
 \nonumber
 \\ & & \mbox{}
 + u_{tz}(u_{xy} + u_{yz}) - u_{ty}(u_{xz} + u_{zz}).
 \label{redrec3}
\end{eqnarray}
With the aid of (\ref{Ia3}), the equation (\ref{redrec3}) takes
the form
\begin{equation}\label{3_2}
  u_{tx} + u_{tz} = u_{tx}u_{tz} - u_{tt}u_{xz} + u_{tz}u_{yz} -
  u_{ty}u_{zz} - \varepsilon .
\end{equation}
After the Legendre transformation
\begin{equation}\label{LegredIa3}
  p = u_x,\quad q = u_z,\quad v(p,q,t,y) = u - xu_x -
zu_z,\quad x = - v_p,\quad z = - v_q
\end{equation}
the equations (\ref{redrec2}) and (\ref{3_2}) take the form
\begin{eqnarray}
 & & v_{pq} = v_{qq} + v_{tq} - v_{py} ,
 \label{leg2}
 \\ & & v_{pq}(v_{tq} - \varepsilon v_{pq} + v_{tt}) = v_{pp}(- \varepsilon v_{qq} + v_{tq} +
 v_{ty}),
 \label{leg3}
\end{eqnarray}
where the equation (\ref{leg2}) was used in the Legendre transform
of (\ref{3_2}) to arrive at (\ref{leg3}). The equation
(\ref{leg3}) can be set into a linear form
\begin{eqnarray}
 & & \lambda v_{pq} = v_{tq}  - \varepsilon v_{qq} +
 v_{ty},
 \label{leg3_1}
 \\ & & \lambda v_{pp} = v_{tq} - \varepsilon v_{pq} + v_{tt}
 \label{leg3_2}
\end{eqnarray}
by introducing an extra unknown $\lambda$ depending on all the
variables. Solving algebraically the system of the three linear
equations (\ref{leg2}), (\ref{leg3_1}) and (\ref{leg3_2}) with
respect to the principal derivatives $v_{ty}$, $v_{tq}$ and
$v_{py}$ in terms of the remaining parametric derivatives in the
form
\begin{eqnarray}
& & v_{ty} = \varepsilon (v_{qq} - v_{pq}) + \lambda (v_{pq} -
v_{pp}) + v_{tt} ,
 \label{5}
 \\ & & v_{tq} = \varepsilon v_{pq} + \lambda v_{pp} - v_{tt} ,
 \label{7}
 \\ & & v_{py} = (\varepsilon - 1)v_{pq} + \lambda v_{pp} + v_{qq}
 - v_{tt} ,
 \label{8}
\end{eqnarray}
we easily check that all cross derivatives of the left-hand sides
coincide as a consequence of these equations, so that this system
of PDEs does not have nontrivial integrability conditions. The
mixed heavenly equation (\ref{Ia3}) after the Legendre
transformation (\ref{LegredIa3}) becomes
\begin{equation}\label{legmix}
  v_{tq}v_{py} - v_{pq}v_{ty} + v_{tt}v_{qq} - v_{tq}^2 +
  \varepsilon (v_{pp}v_{qq} - v_{pq}^2) = 0.
\end{equation}
The Legendre transformed mixed heavenly equation (\ref{legmix})
obviously constitutes another particular case of our general
equation (\ref{Ffinal}), up to a change of notation of the
dependent and independent variables. The linear equations
(\ref{5}), (\ref{7}), (\ref{8}) together with (\ref{legmix}) imply
that $\lambda = - \varepsilon$ as far as $v_{pp}v_{qq} - v_{pq}^2
\ne 0$. Then the Legendre transformed mixed heavenly equation
(\ref{legmix}) becomes an algebraic consequence of these three
linear PDEs with constant coefficients.

In the case when $\varepsilon = -1$ and hence $\lambda = 1$, the
equations (\ref{leg2}) and (\ref{leg3_1}) imply
\begin{equation}\label{conseq}
  v_{ty} + v_{py} = 0 ,
\end{equation}
which can be integrated to yield the linear first-order equation
\begin{equation}\label{1storder}
  v_t + v_p = C(t,p,q) .
\end{equation}
This obviously leads to dependence of $v$ on the characteristic
combination $t - p$ and thus determines invariant solutions. In
this case we have a symmetry reduction and no lift to noninvariant
solutions.

In the case when $\varepsilon = 1$ and hence $\lambda = -1$, the
equations (\ref{leg2}), (\ref{leg3_1}) and (\ref{leg3_2}) do not
imply any linear first-order consequences, so there is no symmetry
reduction of the number of variables in this case and invariant
solutions generically do not arise. Under the change of variables
$(q,p,t)\mapsto (q,\eta = p+t,\xi = p-t)$, the equation (\ref{7})
takes the form of the linear reduced equation (\ref{linred}) but
written in the new variables $\eta,\,\xi$ and $q$
\begin{equation}\label{new813}
  v_{\eta\eta} + v_{\xi\xi} - v_{\xi q} = 0
\end{equation}
and containing the fourth argument $y$ of the unknown $v$ as a
parameter. Any solution of this linear equation depends on the
three variables $\eta,\,\xi$ and $q$ with the ``constants'' of
integration depending on the fourth variable $y$. Certain
appropriate linearly independent combinations of two other
equations (\ref{5}) and (\ref{8}), with the use of (\ref{new813}),
in the new variables take the form
\begin{equation}\label{new812}
v_{\xi q} - v_{\eta q} + v_{\xi y} = 0,
\end{equation}
\begin{equation}\label{new814}
v_{\xi q} + v_{\eta q} - v_{qq} + v_{\eta y} = 0.
\end{equation}
These two equations determine the $y$-dependence of the
``constants'' of integration in the solution of (\ref{new813}) and
hence we obtain the lift of invariant solutions of the Legendre
transformed mixed heavenly equation (\ref{legmix}) to noninvariant
solutions of this equation.

It is easy to obtain an infinite set of exact solutions to linear
equations with constant coefficients. Indeed, we can try the
exponential dependence of $v$ on $\eta$
\[v = \exp{\big(a\eta + b(y)\big)} f(\xi,q,y) ,\]
so that (\ref{new813}) becomes
\begin{equation}\label{feq}
  f_{\xi\xi} - f_{\xi q} + a^2 f = 0.
\end{equation}
For the solution of this equation we can try the following ansatz
\begin{equation}\label{f}
  f = A\cos{(\alpha\xi + \beta q + \theta(y))} + B\sin{(\alpha\xi + \beta q +
  \theta(y))}.
\end{equation}
The expression (\ref{f}) satisfies (\ref{feq}) only if $a =
\pm\alpha\sqrt{\alpha - \beta}$, so that
\begin{eqnarray}\label{v_xiqy}
 & & v = \exp{\Big(\pm\alpha\sqrt{\alpha - \beta}\,\eta +
b(y)\Big)} \nonumber
 \\ & & \times \Big[A\cos{(\alpha\xi + \beta q + \theta(y))}
+ B\sin{(\alpha\xi + \beta q + \theta(y))}\Big].
\end{eqnarray}
The expression (\ref{v_xiqy}) satisfies (\ref{new812}) if
\[\theta(y) = - \beta y,\quad b(y) = \pm\beta\sqrt{\frac{\alpha - \beta}{\alpha}}\, y\]
so that (\ref{v_xiqy}) finally becomes
\begin{eqnarray}
& & v = \exp{\left(\pm\sqrt{\alpha(\alpha-\beta)} \left(\eta +
\frac{\beta}{\alpha}y\right)\right)} \nonumber
\\ & & \times\Big\{ A\cos{[\alpha\xi +
 \beta(q - y)]} + B\sin{[\alpha\xi + \beta (q - y)]} \Big\}.
 \label{sol}
\end{eqnarray}
Surprisingly enough, this expression satisfies identically the
third equation (\ref{new814}), though it is  linearly independent
of the other two equations.

Any linear combination of the solutions of the form (\ref{sol}) is
again a solution of the three linear equations (\ref{new813})
(\ref{new812}) and (\ref{new814}) and hence, with $\eta =
p+t,\;\xi = p-t$, it will satisfy the nonlinear Legendre
transformed mixed heavenly equation (\ref{legmix}) at $\varepsilon
= 1$, since it is a consequence of these linear equations. In the
case of a discrete spectrum, we can choose, for example, the
following linear combination
\begin{eqnarray}
 & & v = \sum\limits_i
\exp{\left(\pm\sqrt{\alpha_i(\alpha_i-\beta_i)} \left(\eta +
\frac{\beta_i}{\alpha_i}\,y\right)\right)}
\Big\{A_i\cos{\big[\alpha_i\xi + \beta_i(q - y)\big]} \nonumber
\\ & & \mbox{} + B_i \sin{\big[\alpha_i\xi + \beta_i (q - y)\big]}\Big\} ,
 \label{fullsol}
\end{eqnarray}
where $\alpha_i$, $\beta_i$, $A_i$ and $B_i$ stand for arbitrary
constants. This is an example of a solution to (\ref{legmix}),
which is obviously noninvariant because it clearly depends on four
independent combinations of the variables $\eta,\,\xi, q$ and $y$.
For the case of a continuous spectrum, the sum in (\ref{fullsol})
should be replaced by an integral.

There is also a class of polynomial solutions. We start with the
ansatz
\begin{equation}\label{poly}
v = A(\eta,\xi,y) \frac{q^2}{2} + B(\eta,\xi,y) q + C(\eta,\xi,y).
\end{equation}
The expression (\ref{poly}) will satisfy linear equations
(\ref{new813}) (\ref{new812}) and (\ref{new814}) if the
coefficients have the form
\begin{eqnarray}
 & & A(\eta,\xi,y) = 3\Big[4g(\eta^2 - \xi^2) + 2h\eta\xi +
ky^2\Big],\nonumber
\\ & & B(\eta,\xi,y) = 3 \Big\{\Big[(4g + h)(\xi^2 - \eta^2) + 2(4g - h)\eta\xi\Big]y \nonumber
 \\ & & \mbox{} + h\eta\xi^2 - 4g\eta^2\xi + \mu(\xi^2 - \eta^2)\Big\},
 \label{ABC}
 \\ & & C(\eta,\xi,y) = k\eta y^3 + 3\Big[h(\eta^2 - \xi^2) -
 8g\eta\xi\Big]y^2 + f(\xi\eta^3 - \eta\xi^3) + (h\eta + \mu)\xi^3
 \nonumber
\\ & & \mbox{} - g\eta^4 + \Big[h\xi^3 + 8g\eta^3 + 12g\eta^2\xi - 3(4g + h)\eta\xi^2 +
3\mu (\eta^2 - \xi^2) - 6\mu\eta\xi\Big]y , \nonumber
\end{eqnarray}
where $f$, $g$, $h$, $k$ and $\mu$ are arbitrary constants.

For solutions independent of $\eta$ we obtain, for example, $v =
(\xi + q - y)^4$ and, since all the three equations are linear,
the sum of this solution and (\ref{ABC}) is again a solution, so
that
\begin{equation}\label{sum}
v = A(\eta,\xi,y) \frac{q^2}{2} + B(\eta,\xi,y) q + C(\eta,\xi,y)
+ D (\xi + q - y)^4 ,
\end{equation}
with $A$, $B$ and $C$ defined by (\ref{ABC}) and constant $D$,
will satisfy the nonlinear Legendre transformed mixed heavenly
equation (\ref{legmix}). This solution generically depends on all
the four variables and hence it is a noninvariant solution, i.e.
it does not admit Lie symmetries. More general polynomial
solutions can easily be constructed. The sum of the exponential
solution (\ref{fullsol}) and a polynomial solution again satisfies
(\ref{legmix}).

\section{Ricci-flat metrics governed by the mixed\\ heavenly equation and its Legendre\\ transform}
 \setcounter{equation}{0}
 \label{sec:metrics}

In section \ref{sec:lift}, we have obtained noninvariant solutions
(\ref{fullsol}) and (\ref{sum}) of the Legendre transformed mixed
heavenly equation (\ref{legmix}). In order to get the
corresponding solution of the mixed heavenly equation (\ref{Ia3}),
we had to perform the Legendre transformation of solutions
(\ref{fullsol}) and (\ref{sum}), inverse to (\ref{LegredIa3}),
which is quite a difficult problem.

Instead, we shall proceed, as we did before in
\cite{mns,mnsh,lift}, by taking into account that, similar to the
complex Monge-Amp\`ere equation and second heavenly equation of
Pleba\~nski, the mixed heavenly equation determines a potential
that governs Ricci-flat metrics in the self-dual gravity. If we
are interested only in such metrics as our final result, then
instead of performing the inverse Legendre transformation of our
solution, we make the direct Legendre transformation
(\ref{LegredIa3}) of the metric related to the mixed heavenly
equation. Then our solutions (\ref{fullsol}) and (\ref{sum}) of
the Legendre transformed mixed heavenly equation (\ref{legmix}) at
$\varepsilon = 1 $, or any other its solutions determined by the
linear equations (\ref{new812}), (\ref{new813}) and
(\ref{new814}), will yield a potential governing the Legendre
transformed \textit{mixed heavenly metric}.

In order to obtain Ricci-flat metrics related to the mixed
heavenly equation, we start with Husain's heavenly metric and then
use the relation between the Husain's equation and mixed heavenly
equation. Husain's heavenly metric has the form \cite{husain}
\begin{equation}\label{husmetr}
  ds^2 = 2\left(\omega_t dt + \omega_p dp + \frac{(\omega_t^2 +
  \omega_p^2)}{\Delta_{tp}}\right),
\end{equation}
where
\[\omega_t = \Lambda_{ty}dy + \Lambda_{tz}dz,\quad \omega_p = \Lambda_{py}dy +
\Lambda_{pz}dz ,\quad \Delta_{tp} = \Lambda_{ty}\Lambda_{pz} -
\Lambda_{tz}\Lambda_{py} \]
 with $\Lambda(t,p,y,z)$ satisfying the Husain's equation
\begin{equation}\label{huseq}
  \Lambda_{tt} + \Lambda_{pp} + \Lambda_{tz}\Lambda_{py} -
  \Lambda_{ty}\Lambda_{pz} = 0.
\end{equation}
By the one-dimensional Legendre transformation
\begin{equation}\label{hus_mix}
  \Lambda = u - xu_x,\quad p = -u_x, \qquad x = \Lambda_p,\quad u =
  \Lambda- p\Lambda_p ,
\end{equation}
where the inverse transformation is also given, the Husain's
equation (\ref{huseq}) is mapped into the mixed heavenly equation
with $\varepsilon = + 1$
\begin{equation}\label{mix}
u_{ty}u_{xz} - u_{tz}u_{xy} + u_{tt}u_{xx}-u_{tx}^2 = 1
\end{equation}
for the unknown $u(t,x,y,z)$. Performing the transformation
(\ref{hus_mix}) of the Husain's metric (\ref{husmetr}), we obtain
the metric governed by equation (\ref{mix})
\begin{equation}\label{mixmet}
  ds^2 = 2\left\{\omega_t dt + \omega_x dx
  + \frac{1}{u_{xx}\Delta}\left[(u_{xx}\omega_t - u_{tx}\omega_x)^2 + (\Delta +
  1)\omega_x^2\right]\right\} ,
\end{equation}
where
\begin{equation}\label{def}
  \omega_t = u_{ty} dy + u_{tz} dz,\quad \omega_x = u_{xy} dy + u_{xz} dz,
\quad \Delta = u_{tz}u_{xy} - u_{ty}u_{xz}.
\end{equation}
By using a REDUCE program, it has been checked that the metric
(\ref{mixmet}) is Ricci-flat as a consequence of equation
(\ref{mix}).

The asymmetry of the metric (\ref{mixmet}) in variables $t$ and
$x$ is caused by the Legendre transformation (\ref{hus_mix})
between $p$ and $x$, which leaves $t$ untransformed. To amend this
lack of symmetry, we symmetrize the metric (\ref{mixmet}) in $t
\leftrightarrow x$ and $y \leftrightarrow z$ and then introduce
$t\pm x$ and $y\pm z$ as new coordinates, which we call again
$t,x,y,z$. The resulting \textit{mixed heavenly metric} has the
form
\begin{equation}\label{mixheavmet}
ds^2 = 2\left\{\omega_t dt + \omega_x dx
  + \frac{1}{\Delta}\left(u_{xx}\omega_t^2 - 2u_{tx}\omega_t\omega_x  +
  u_{tt}\omega_x^2\right)\right\}.
\end{equation}
This metric is also Ricci-flat, provided that the potential $u$
satisfies the mixed heavenly equation (\ref{Ia3}) \textit{for both
signs of $\varepsilon$}. Now we apply the Legendre transformation
(\ref{LegredIa3}) to the mixed heavenly metric (\ref{mixheavmet})
with the result
\begin{eqnarray}
  ds^2 = 2\left\{ \frac{1}{\delta}\,(v_{tt} + \varepsilon v_{pp})
(v_{qt}dt + v_{qp}dp + v_{qq}dq)^2\right.\nonumber
 \\ \left. \mbox{} + (v_{qt}dt + v_{qp}dp +
v_{qq}dq)\left[- dq + \frac{2}{\delta}(v_{tq}v_{ty} + \varepsilon
v_{pq}v_{py})dy\right]\right.\nonumber
\\ \left. \mbox{} + \left[v_{yt}dt + v_{yp}dp +
\frac{v_{qq}}{\delta} (v_{yt}^2 + \varepsilon v_{yp}^2) dy \right]
dy \right\},
 \label{legmixmet}
\end{eqnarray}
where $\delta = v_{ty}v_{pq} - v_{tq}v_{py}$ and the metric
potential $v(t,p,q,y)$ should satisfy the Legendre transformed
mixed heavenly equation (\ref{legmix}). With the latter condition
satisfied, by using REDUCE we have checked that the metric
(\ref{legmixmet}) is Ricci-flat and calculated the Riemann
curvature tensor components for an arbitrary $v$ satisfying
(\ref{legmix}). The expressions for these components are too
lengthy to be presented for publication. However, the denominators
of the Riemann tensor components are simple, so that possible
singularities of the curvature tensor either coincide with the
singularities of the metric (\ref{legmixmet}), being at $\delta
\equiv v_{ty}v_{pq} - v_{tq}v_{py} = 0$, or are located at $v_{qq}
= 0$, for $v$ being a linear function of $q$. For the polynomial
solution (\ref{sum}) the condition $\delta = 0$ could be satisfied
only if all the essential coefficients in (\ref{sum}) vanished: $h
= g = \mu = D = 0$, which would contradict the non-invariance of
this solution. The only singularity of the metric corresponding to
(\ref{sum}) is located at infinity.

As it was shown in section \ref{sec:lift}, we can use any solution
of the three linear equations (\ref{new813}) (\ref{new812}) and
(\ref{new814}), which imply (\ref{legmix}) at $\varepsilon = 1$ as
their algebraic consequence. In particular, we can use the
noninvariant solutions (\ref{fullsol}) and (\ref{sum}) for $v$ in
the metric (\ref{legmixmet}). For noninvariant solutions, there
will be no symmetry reduction, so that $v$ will depend on all the
four independent variables, which is a necessary (and often
sufficient) condition for the metric components in
(\ref{legmixmet}) to depend also on all the four independent
variables. For the exponential solutions to the complex
Monge-Amp\`ere equation and to the second heavenly equation, which
are similar to (\ref{fullsol}), we have proved in \cite{mnsh} that
the corresponding K\"ahler metric and the second heavenly metric
admit no Killing vectors. Similarly, for the solution
(\ref{fullsol}) we also expect that the Legendre transformed mixed
heavenly metric (\ref{legmixmet}) will admit no Killing vectors
and hence no symmetry reduction in the number of independent
variables will occur.

\section{Conclusion}

 In the theory of gravitational instantons, heavenly metrics with
no Killing vectors (no continuous symmetries) can only be
generated by noninvariant solutions of $CMA$. Therefore, we are
faced with the problem of obtaining noninvariant solutions of
partial differential equations. Partner symmetries proved to be an
appropriate tool for solving such a problem because noninvariant
solutions can be obtained as solutions invariant with respect to a
certain nonlocal symmetry closely related to partner symmetries.
Thus, the existence of partner symmetries for a given PDE is
necessary to apply this method. In this paper, we have obtained a
general form of the scalar second-order PDE in four variables,
containing only second derivatives of the unknown, that possesses
partner symmetries. Using point and Legendre transformations, we
have transformed this general equation to different simplest
canonical forms and so presented a classification of inequivalent
equations which admit partner symmetries, together with recursion
relations for symmetries. Among these equations we find the
well-known first and second heavenly equations of Pleba\~nski and
two other nonlinear equations which we have called \textit{mixed
heavenly equation} and \textit{asymmetric heavenly equation}. The
mixed heavenly equation is related by a partial Legendre
transformation to Husain's heavenly equation arising in the chiral
model approach to self-dual gravity. A particular case of the
asymmetric heavenly equation is the evolution form of the second
heavenly equation.

We ignored here all the cases when the canonical equation
explicitly contains only three variables. We leave for the future
a classification of PDEs with three variables, that admit partner
symmetries.

As an example of application of partner symmetries, we have shown
how to construct noninvariant solutions of the Legendre
transformed mixed heavenly equation. By applying Legendre
transformation in two variables, the latter equation and
differential constraints, that are obtained from recursion
relations for partner symmetries, have been transformed to a set
of three linear equations with constant coefficients, that imply
the Legendre transformed mixed heavenly equation as their
algebraic consequence. One of these equations involves only three
variables and formally coincides with a certain reduced equation,
which determines invariant solutions of the Legendre transformed
mixed heavenly equation, but written in new variables and
containing also the fourth variable as a parameter. Two other
equations, involving all the four variables, provide a lift from
invariant to noninvariant solutions of the Legendre transformed
mixed heavenly equation. We have obtained Ricci-flat metrics
governed by the mixed heavenly equation and the Legendre
transformed mixed heavenly equation. Using any noninvariant
solution of the three linear PDEs, we satisfy the necessary
condition of arriving at Ricci-flat metrics with metric components
depending on all four independent variables. Such metrics will
admit no continuous symmetries and no Killing vectors.

Thus, we conclude that, for a scalar second-order PDE with four
independent variables, the existence of partner symmetries happens
to be a characteristic feature of the equations that describe
self-dual gravity in different variables. The partner symmetries
provide a tool for obtaining noninvariant solutions of these
equations and Ricci-flat self-dual metrics with no Killing
vectors.

\section*{Acknowledgements}

One of the authors (MBS) is grateful to George Bluman for
illuminating discussions. The research of MBS is partly supported
by the research grant from Bogazici University Scientific Research
Fund, research project No. 07B301.

\end{document}